\documentclass[jgrga]{agutex}

\usepackage{color}
\usepackage{graphicx}
\usepackage{dcolumn}
\usepackage{bm}

\graphicspath{{figs/}}

\authorrunninghead{GALLANA ET AL.}
\titlerunninghead{VOYAGER 2 SOLAR WIND SPECTRAL ANALYSIS}

\begin{document}

\title{Voyager 2 solar plasma and magnetic field spectral analysis for intermediate data sparsity}



\authors{Luca Gallana,\altaffilmark{1},
Federico Fraternale,\altaffilmark{1}, 
Michele Iovieno \altaffilmark{1},
Sophie M. Fosson \altaffilmark{2},
Enrico Magli \altaffilmark{2},
Merav Opher  \altaffilmark{3},
John D. Richardson \altaffilmark{4},
Daniela Tordella \altaffilmark{1}}

\altaffiltext{1}{Dipartimento di Ingegneria Meccanica e Aerospaziale,
Politecnico di Torino, Torino, Italy}
\altaffiltext{2}{Dipartimento di Elettronica e Telecomunicazioni,
Politecnico di Torino, Torino, Italy}
\altaffiltext{3}{Astronomy Department, Boston University, Boston, MA, USA}
\altaffiltext{4}{Kavli Institute for Astrophysics and Space Research, Massachusetts Institute of Technology (MIT), Cambridge, MA, USA}

\begin{abstract}
The Voyager probes are the furthest, still active, spacecraft ever launched from Earth. During their 38-year trip, they have collected data regarding solar wind properties (such as the plasma velocity and magnetic field intensity). Unfortunately, a complete time evolution of the measured physical quantities is not available. The time series contains many gaps which increase in frequency and duration at larger distances.
The aim of this work is to perform a spectral and statistical analysis of the solar wind plasma velocity and magnetic field using Voyager 2 data measured in 1979, when the gaps/signal ratio is of order of unity. This analysis is achieved using four different data reconstruction techniques: averages on linearly interpolated subsets, correlation of linearly interpolated data, compressed sensing spectral estimation, and maximum likelihood data reconstruction.
With five frequency decades, the spectra we obtained have the largest frequency range ever computed at 5 astronomical units from the Sun; spectral exponents have been determined for all the components of the velocity and magnetic field fluctuations. Void analysis is also useful in recovering other spectral properties such as integral scales (see for instance Table 4) and, if the confidence level of the measurements is sufficiently high, the decay variation in the small scale range due, for instance, to dissipative effects.

\bf{Key points}: 

- solar wind data void reconstruction with four different methods

- enhanced fluctuation  power spectral analysis

- over five decades of frequency spectral range
\end{abstract}




\begin{article}


\section{Introduction}
\label{cha:intr}


The solar wind fills the heliosphere from the Sun to the termination shock with a supersonic flow of magnetized plasma. This flow is time-dependent on all scales and expands with distance. The flow has fluctuations on a broad range of scales and frequencies. These fluctuations are not just convected outward but show energy
cascades between the different scales. The solar wind turbulence phenomenology has been comprehensively reviewed  by \cite{tu1995,bruno2013}. 

Most studies of solar wind turbulence use data from near-Earth, with spacecraft in the ecliptic near 1 AU, see \cite{tu1995}. Recent studies of the solar wind near 1 AU found the fluctuations in magnetic field are fit by power laws with exponents of -5/3 while those of velocity often show exponents of -3/2 \cite{podesta2007}. The Ulysses spacecraft provided the first observations of turbulence near the solar polar regions \cite{horbury2001}; hourly-average Ulysses data show that the velocity power law exponent evolves toward -5/3 with distance from the Sun, and that spectra at 1 AU are far from the asymptotic state \citep{roberts2010}. In order to understand the evolution of the solar wind and its properties, it is necessary to analyze data at larger radial distances. However, data gaps typically increase with  distance and make the spectral analysis challenging.


In this paper Voyager 2 (V2) plasma and magnetic field data from near 5 AU are used to study the structure of turbulence in the solar wind. Voyager 2 was launched in August 23, 1977 and reached a distance of 5 AU in the first half of 1979 (just before the Jupiter fly-by. V2 closest approach to Jupiter was on July 9). We use data from January 1 to June 29, 1979 (DOY 1 - 180) . The Voyager plasma experiment observes plasma currents in the energy/charge range 10 - 5950 eV /q using four modulated-grid Faraday cup detectors \citep{bridge1977}. The observed currents are fit to convected isotropic proton Maxwellian distributions to derive the parameters (velocity, density, and temperature) used in this work. Magnetic field and plasma data are from the COHOWeb repository (http://omniweb.gsfc.nasa.gov/coho/) of the Space Physics Data 
Facility.
In 1979 data gaps are due mainly to tracking gaps; some smaller gaps are due to interference from other instruments. As a consequence, datasets from Voyager 2 are lacunous and irregularly distributed. In order to perform spectral analysis, methods for signal reconstruction of missing data must be implemented.

In section two we present the physical behavior of the solar wind near 5 AU through statistical analysis and plasma parameters.
Section three gives an overview of the signal context we work with and of the reconstruction techniques used. In section four we show and discuss the spectra analysis performed about kinetic energy, density and thermal speed.  Conclusions and future development follow in section five. Supplementary Information provides practical details on the software used carry out the gapped data analysis.
\section{1979 DAY 1-180 Voyager 2 data. Probability density function and intermittency. Limitation toward spectral analysis}
\label{cha:data}
%


	The dataset consists of vector plasma velocity and magnetic field data from 01/01/1979 00:00 GMT to 06/29/1979 19:00 GMT, a period of about 180 days.
	In 1979 the plasma speed and direction were sampled each 96 s, while for magnetic field the resolution of the data we use is 48 s (the actual sampling frequency is  higher than 0.1 Hz).
	
\begin{figure}
\centering
	\includegraphics[width=.45\textwidth]{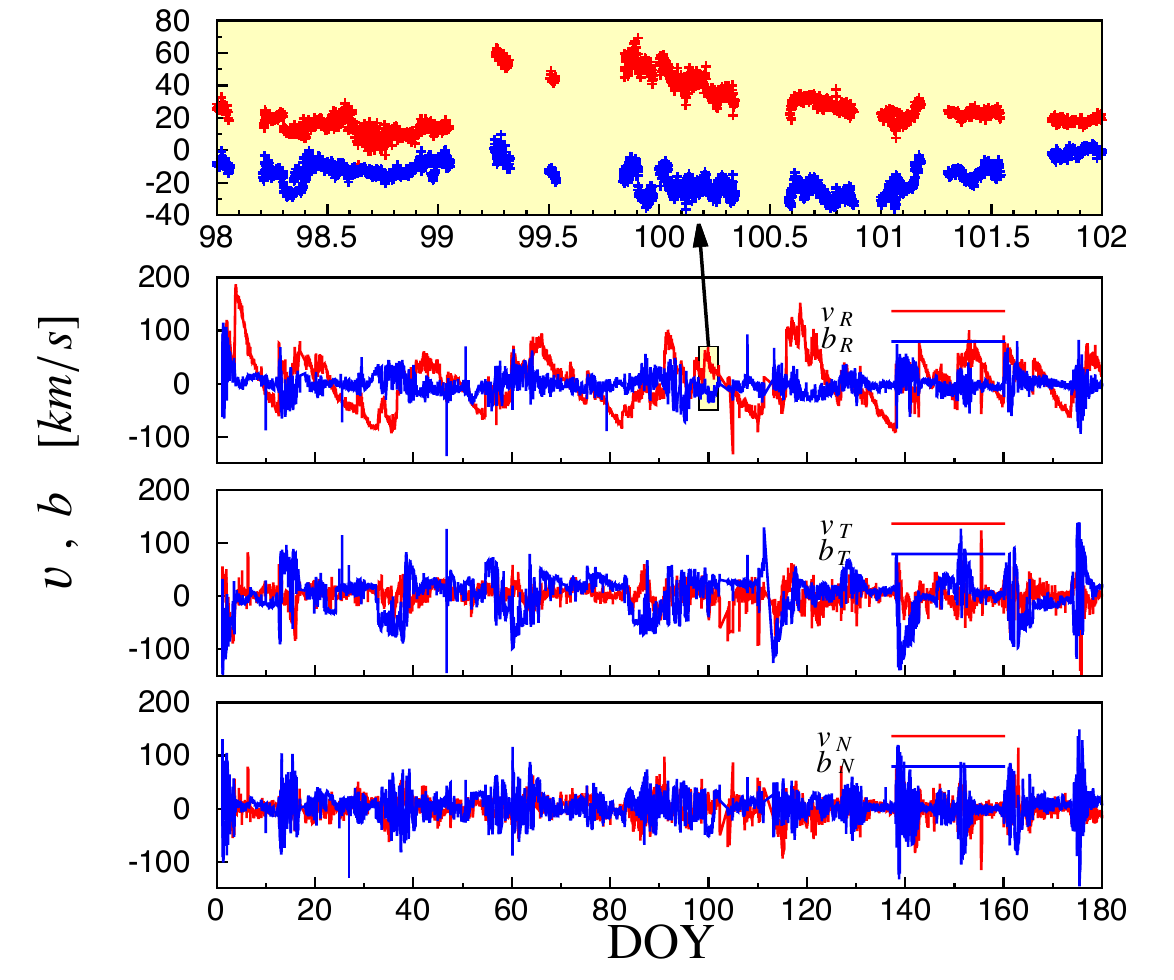}
	\caption{Plasma velocities (red lines) and magnetic fields (blue lines) recorded by Voyager 2 in the first 180 days of 1979. RTN Heliographic reference frame is used. The magnetic field is represented  using Alfv\'en units in order to compare the two different datasets. In the top panel 4 days period is magnified to show  the data gaps.}
	\label{fig:data_zoom}
\end{figure}

	In this period the solar wind was mainly slow, with a mean velocity $V_{SW}$ equal to $454\ km/s$ and relatively high levels of fluctuation, see table \ref{parameter}. The slow wind is characterized by magnetic fluctuations generally comparable to the mean field, with values of the magnetic energy ($E_m= 2.84\cdot10^3\ km^2/s^2$) greater than the kinetic energy ($E_k=2.45\cdot10^3\ km^2/s^2$) \citep{marsch1990a, mccomas1998}; the average Alfv\'en ratio $r_A$ in the period is about $0.862$.  In the same table other characteristic plasma frequencies and scales are reported.
%
A plot of the data  is shown in figure \ref{fig:data_zoom}, where the fluctuations of the components of plasma velocity and Alfv\'en velocity are represented using the RTN Heliographic reference system. The RTN system is centered at the spacecraft, the R (radial) axis is directed radially away from the Sun through the spacecraft. The T (tangential) axis is the cross product of the Sun's spin vector (North directed) and the R axis, i.e. the T axis is parallel to the solar equatorial plane and is positive in the direction of planetary rotation around the Sun. The N (normal) axis completes the right handed set). The top panel of the figure magnifies 4 days of data to show the typical data gap distribution.


	\begin{table}
	
		\caption{\label{parameter}Reference parameters for the considered period. All the  quantities are averages (or integral) over the whole period of 180 days.}
		\centering
				\begin{tabular}{@{}llll}
					\hline
					\multicolumn{2}{c}{Parameter} &  \multicolumn{2}{c}{Value}                      \\
					\hline
					\boldmath $v_{SW}$        & Plasma velocity                        & $4.54\cdot10^2$  & km/s         \\
					\boldmath $V_{A}$           & Alfv\'en velocity                      & $4.94\cdot10^1$            & km/s         \\
					\boldmath $E_k$             & Kinetic energy                       & $2.45\cdot10^3$  & km$^2$/s$^2$ \\
					\boldmath $E_m$             & Magnetic energy                      & $2.84\cdot10^3$  & km$^2$/s$^2$ \\
					\boldmath $E$               & Total energy                         & $5.29\cdot10^7$  & km$^2$/s$^2$ \\
					\boldmath $H_c$             & Cross helicity                       & 15.8              & km$^2$/s$^2$ \\
					\boldmath $H_m$             & Magnetic helicity                    & $2.10\cdot10^6$ & nT$^2$km     \\
					\boldmath $ n_{i}$          & Numerical density                    & 0.23             & cm$^{-3}$    \\
					\boldmath $ E_T$            & Thermal energy                       & 2.29             & eV           \\
					\boldmath $ T$              & Temperature                          & $2.70\cdot10^4$         & K            \\
					\boldmath $\beta_{p}$       & Plasma beta                          & 0.22            &  \\
					\boldmath $c_{s}$           & Sound speed                          & $1.93\cdot10^1$          & km/s         \\
					\boldmath $f_{ci}$          & Ions Larmor frequency                & 0.02            & Hz           \\
					\boldmath $f_{pi}$          & Ions plasma frequency                & 0.10              & kHz           \\
					\boldmath $f*$              & Convective Larmor frequency          & 0.44             & Hz           \\
					\boldmath $r_{ci}$          & Larmor radius                        & $4.29\cdot10^3$             & km           \\
					\boldmath $r_{i}$           & Ion inertial radius                 & $1.58\cdot 10^2$              & km           \\
					\boldmath $\lambda_{D}$     & Debye length                         & 5.5              & m            \\
					\hline
				\end{tabular}
		\end{table}

		\begin{table}
		\centering
		\caption{\textbf{Intermittency and anisotropy for the solar wind}. Mean values and first three moments for the velocity and magnetic fields components and for their fluctuation modules. $\mu$ is the mean value, $\sigma^2$ the variance, $Sk$ the skewness and $Ku$ the kurtosis. 
		Velocity units are $km/s$ and $km^2/s^2$ for the mean and variance respectively, while for magnetic field the units are  $nT$ and $nT^2$. Skewness and kurtosis are dimensionless. The modules of the fluctuations are normalized on the variance, see equation \ref{eq:norm}, in order to be able to compare them with a chi-square distribution (standard 3-component chi-square distribution has mean 3, variance 6, skewness 1.63 and kurtosis 7). }
		\label{tab:pdf}
		\begin{tabular}{lrrrrr}
			\hline         
			 \  &  \ \ & $\mu$  &  $\sigma^{2}$& $Sk$ &  $Ku$  \\ \hline
			$v_{R}$& \ \  &   454      &  1893      &  0.43   &  3.41\\ 
			$v_{T}$& \ \  &   3.21     &  252.9     & -0.99   &  7.35\\ 
			$v_{N}$& \ \  &   0.51     &  250.3     & -0.36   &  5.80\\
			$B_{R}$& \ \  &  -0.04     &  0.173     &  0.53   &  6.71\\ 
			$B_{T}$&  \ \  &  0.06     &  0.85      &  -0.72  &  10.2\\ 
			$B_{N}$& \ \  &   0.10     &  0.34      &  -0.24  & 7.65\\
			$|\mathbf{\delta v}|$ & \ \  & 3.00 & 10.47 & 2.40 & 10.27 \\
			$|\mathbf{\delta B}|$ & \ \  & 2.48 & 17.41 & 3.17 & 14.90\\
			\hline
		\end{tabular}	
	\end{table}

\begin{figure*}
	 	\centering
		\includegraphics[width=.80\textwidth]{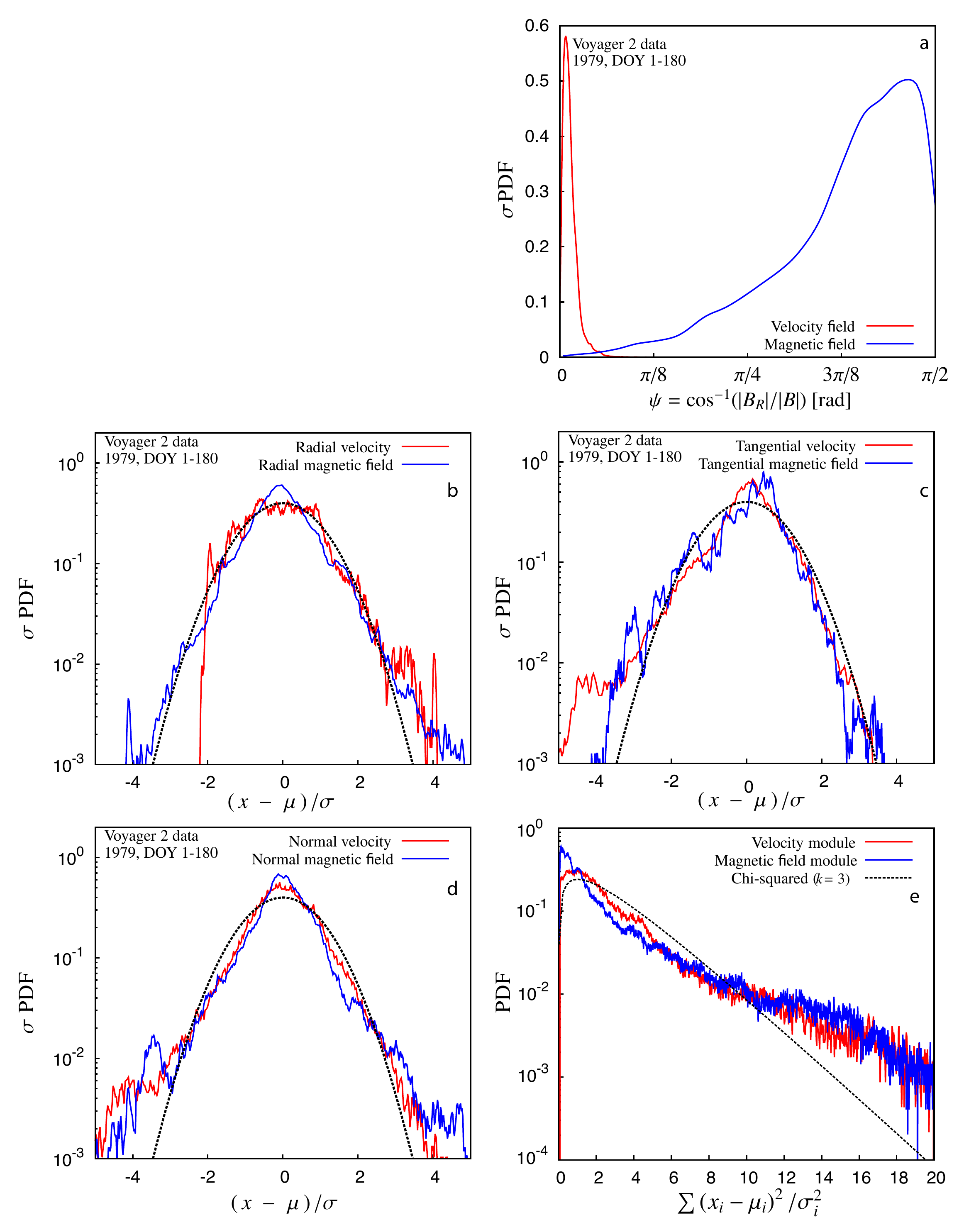}
		\caption{
	 		(\textit{a}) Normalized probability density function of the angle $\psi$ between the radial direction and the local velocity direction (red line) and magnetic direction (blue line),  respectively. The magnetic field is generally tilted  85 to 105 degrees with respect to the the radial direction. (\textit{b}-\textit{c}-\textit{d}) Normalized probability density function of the plasma velocity and magnetic field comparing the components in the radial direction (\textit{a}), the tangential direction (\textit{b}) and normal direction (\textit{c}) and the normalized modules (\textit{d}). For the components (panels \textit{b - d}), the black dashed lines represent a normal gaussian distribution. Differences between the components (and from the reference curve) indicate a non-gaussian, anisotropic behavior. The velocity field presents more asymmetric distributions, while the PDFs of magnetic field deviate most from gaussianity (that means higher kurtosis values). For the normalized modules (panel \textit{e}), the black dashed line represent a chi-square distribution for 3D vectorial variables. The modules are normalized as defined in equation \ref{eq:norm}.  }
	 		 	\label{fig:pdf}
	 \end{figure*}

	The anisotropy of the fields can be determined  by looking at the single components probability density functions (PDFs) in figure \ref{fig:pdf}, panels (\textit{a-c}). Particularly important are the differences of the radial components compared to the tangential and normal ones: a quantification of the anisotropy can be appreciated by comparing the skewness values in table \ref{tab:pdf}.
	The presence of intermittency in the velocity and magnetic fields can be also observed by looking at the PDFs of the modules of the normalized vector fields, shown in figure \ref{fig:pdf} (\textit{d}). The normalized vector fields are given by
	\begin{equation}
	\label{eq:norm}
	|\mathbf{\delta x}|=\sum_{i}^{3}\frac{(x_i-\mu_i)^2}{\sigma^2_i}
	\end{equation}
	where $\mu_i$ is the mean value and $\sigma_i^2$ the variance of the i-th component of the vector field $\mathbf{x}$. The same plot shows a three-component chi-square distribution as a reference. Intermittency occurs over a broad range of scales and seems to be slightly higher in the magnetic field data which has larger skewness and kurtosis (see table \ref{tab:pdf}).

 In order to analyze anisotropic effects from a spectral point of view, it is important to identify the wave-numbers parallel to the magnetic field $k_\parallel$ and normal to it $k_\bot$, as suggested first by \cite{montgomery1987}. We consider the angle $\psi$ between the local vector field and the radial direction, defined as
 \begin{equation}
 	\psi_v=\cos^{-1}\left(\frac{\left|v_r\right|}{\left|\mathbf{v}\right|}\right)\qquad 	\psi_m=\cos^{-1}\left(\frac{\left|B_r\right|}{\left|\mathbf{B}\right|}\right).
 \end{equation}
 The PDFs of these angles are represented in panel \bf{a} of figure \ref{fig:pdf}: while the average plasma velocity is oriented along the radial direction (the mean angle is $0.04\pm0.03$ radiant), moving away from the sun, the magnetic field is characterized by angles close to $\pi/2$ which make it perpendicular to the radial direction as expected for a Parker spiral.

 Since data collected in time can be interpreted as if they were measured along the radial direction (the solar wind is radial, and the probe motion is very low compared with the wind velocity), the perpendicular wave-number can be identified by using the relation

 \begin{equation}
 \label{eq:perp}
 	 f \approx v_{SW} k_R\approx v_{SW} k_\bot  
 \end{equation}

\noindent where $v_{SW}$ is the mean plasma velocity and $k\bot$ are the wave-numbers normal to the mean magnetic field. This approximation can be considered valid since the radial direction is normal to the mean magnetic field, as shown in panel \bf{a} of  figure \ref{fig:pdf}.  

\newpage
	
	\begin{figure*}
	 \centering
		\includegraphics[width=.6\textwidth]{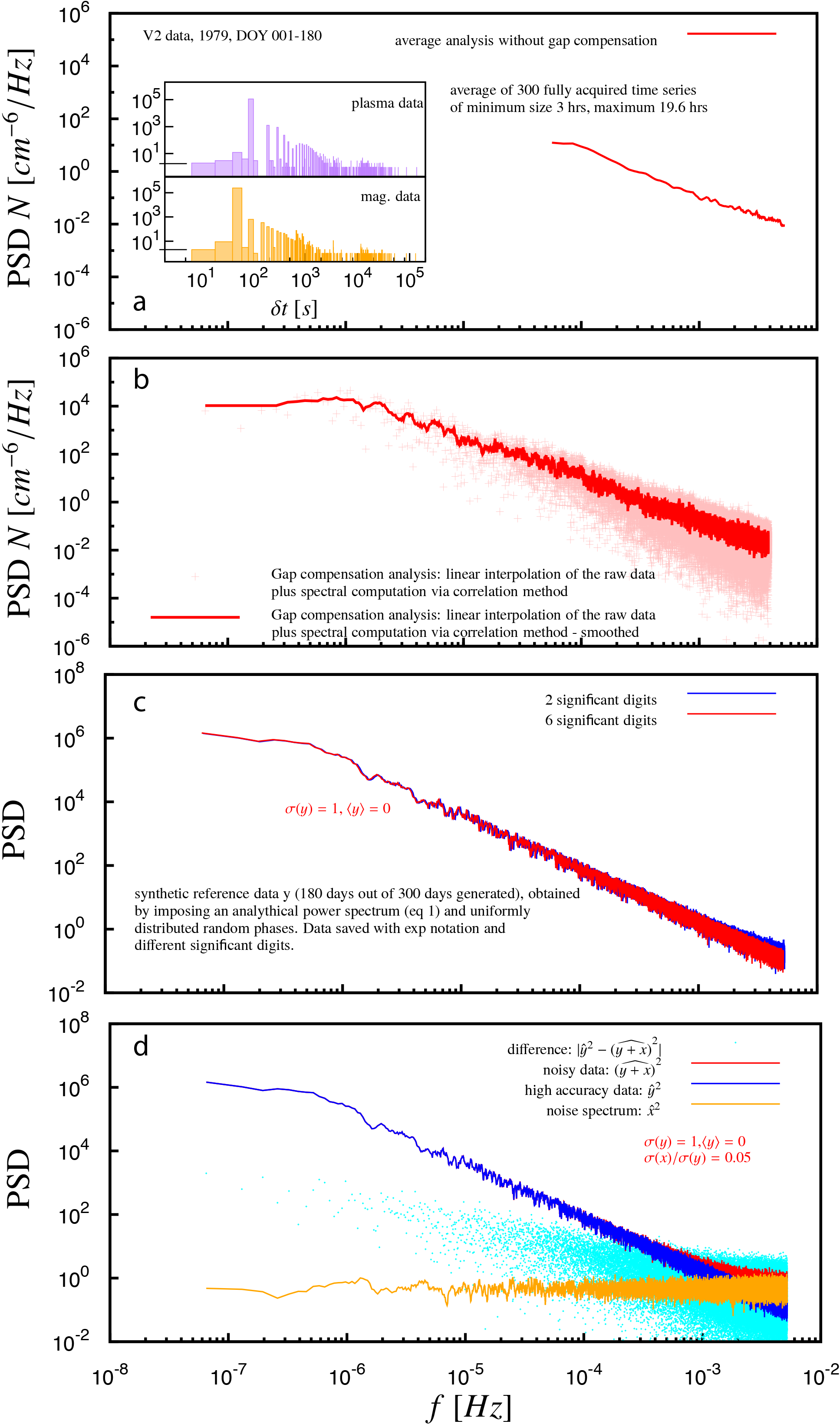}
		\caption{Context of the data sparsity spectral analysis. Panel a: Averaged of density spectrum computed for all time intervals (3 to 19.6 hours) in the first half of 1979 which had no missing data. The inset shows plasma and magnetic field gap distributions. Panel b: the spectra derived with  gap compensation which cover two decades and a half more in frequency range and includes the frequency transition between the energy injection and the interial ranges. This last frequency allows the determination of the system temporal integral scale (see for instance in Table 4 the temporal integral scales deduced in this work for the kinetic energy and the magnetic field). The smoothing used all throughout this work is light (averages over 9  side frequency values) and uniform over all the spectral range to prevent the intriduction of spuriopus results from the data cleaning. Panel c shows the minor role played by the precision of the data which is used once a given level of accuracy is obtained. Panel d: by comparing the blue, red and yellow plots, one can see the curtailment of the maximum observable frequency due to overall confidence level possessed by the measurement and data acquisition chain. The unsmoothed dotted light blue spectrum shows the addition of power spectral energy due to the presence of signal noise.}	 	
	 	\label{fig:generale}
		\end{figure*}

\section{Data reconstruction methods for long term temporal spectra analysis}
\label{cha:recons}
Data sparsity represents one of the major challenges to spectral analysis. Figure \ref{fig:generale} presents the general context of gapped data treatment including details of the role played by measurement accuracy, numerical precision, presence of noise and related energy addition. The gap distribution of the V2 velocity and magnetic field data is shown in the insert of panel \bf{a} of figure \ref{fig:generale}. In the 180-day period analyzed,  $28 \%$ and $24 \%$ of plasma and magnetic field data are missing, respectively. Naming $\delta t$ the time difference between consecutive data points, the biggest gap is $\delta t_{max}= 44.7$ hours, and the total number of points is 115102 for the velocity and density, 248159 for the magnetic field. The longest continuous data subset lasts $Ts=19.5$ hours and it is located at DOY 176 and 168 for plasma and magnetic fields, respectively. In the following, the temporal length of a data gap is named $Tg$. 

In order to test different spectral analysis procedures, two different 180-day reference numerical datasets have been prepared that mimic the V2 data in terms of integral scale and sampling period. The datasets are  called synthetic turbulence 1 and 2 and were prepared by analogy to that of a  scalar field which has a  power spectrum similar to the typical one-dimensional spectrum of an hydrodynamical homogeneous and isotropic turbulent field, see \citet{monin_book}. The two reference fields contain 

\begin{itemize}
\item  \textit{Synt 1} : an energy injection range and an inertial range 
\item  \textit{Synt 2} :  an energy injection range, an inertial range and a dissipative range.   
\end{itemize}
\vspace{10 pt}

In these synthetic reference datasets, the energy injection range follows a power law with exponent in the interval $2 \pm 1.5$, the spectral maximum is placed at a frequency corresponding to one solar day, the inertial range extends over three and a half decades and has a power decay equal to $ -5/3 \pm 1$, and the dissipative range is placed around $5 \cdot 10^{-3}$ Hertz and has a maximum decay of -3. Furthermore, the phases of the harmonic components have been uniformly randomized.

In the following, we show results for an energy injection rate of 2 and a inertial decay rate of -5/3. However,  for all exponents in the above intervals, the same good behaviour is shown  by all the gap compensation methods here presented.
  
\begin{figure*}
 	\centering
 	\includegraphics[width=.80\textwidth]{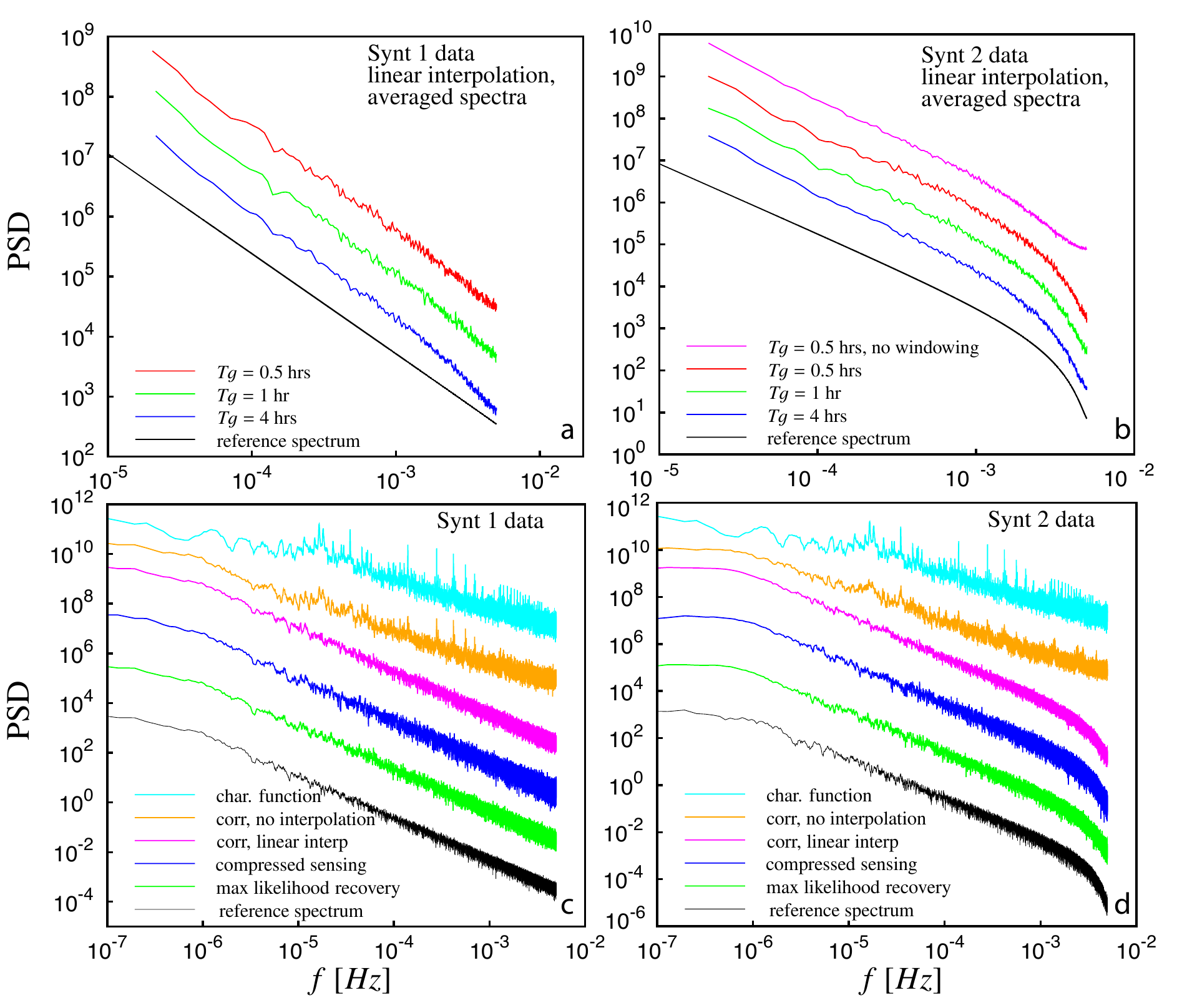}
 	\caption{Validation of spectral analysis techniques for synthetic turbulence lacunous data. A comparison with the original complete sequences (see the black curves) allows us to estimate the error in the spectral slopes. ({\bf a},{\bf b}) Averaged spectra. A direct Fourier transform, with Hann windowing, is performed on linearly interpolated subsets. These segments are selected so that the maximum gap length filled by the interpolation is $Tg$. The linear interpolator has a low-pass effect, evident from panel ({\bf a}) in the high frequency range. Here the error on the spectral index in the range $f\in[10^{-5}-10^{-3}]$ lies between $1.9\%$ ($Tg=0.5$ hrs) and $5.4\%$ ($Tg=4$ hrs). For the \textit{Synt2} data,  windowing is helpful to recover the correct spectral slopes, see the pink curve of panel ({\bf b}). Here the error lies between $0.4\%$  ($Tg=0.5$ hrs) and $2.4\%$ ($Tg=4$ hrs) in the range $f\in[10^{-5}-10^{-3}]$. ({\bf c},{\bf d}) Spectral computation for the full frequency range. Cyan: spectrum of the characteristic function $\phi$. Orange:  spectrum of the two-point correlation function, computed from the gapped sequence. The influence of gaps is evident, the plot shows non-physical peaks and the wrong slope. Correlations of linearly interpolated data, in the whole range, show much better convergence, see the pink curve. Blue: compressed sensing spectral estimation (BPDN formulation). Green: spectrum from maximum likelihood data reconstruction \citep{rybicki1992a}. Black: Fourier transform of the original complete sequences. The discrepancy of the power law exponent is below 2.5\% for the last three methods. To all these spectra a smoothing is applied by averaging neighboring frequencies.  The energy is preserved for all spectra, but they have been shifted for clarity. }
 	\label{fig:method}
 \end{figure*}

These sequences have been made sparse by giving them the same gap distribution as the Voyager 2 data. The spectra of these reference synthetic datasets are represented with black curves in figure \ref{fig:method} which illustrates the behavior  of four techniques for spectral analysis of lacunous data: i) windowed averaged Fourier transforms of linearly interpolated data subsets; ii) Fourier transform of the correlation function;  iii) maximum likelihood recovery by \cite{rybicki1992a}; iv) spectral estimation via compressed sensing (see e.g. \cite{donoho2006, candes2006}).

Results from the first technique are shown in panels ({\bf a},{\bf b}) of figure \ref{fig:method} for different values of $Tg$, which is the maximum size of the gap where data are interpolated. The low-pass behavior of the interpolator results in a steepening of the spectrum, especially in the high frequency range, and becomes much more evident as $Tg$ increases. For \textit{Synt1}, the relative error on the spectral index $\alpha$ lies between $1.9\%$  ($Tg=0.5$ hrs) and $5.4\%$ ($Tg=4$ hrs) in the range $f\in[10^{-5},10^{-3}]$, while in the last frequency decade it increases up to $8\%$ for $Tg=0.5$ hrs. For  \textit{Synt2}, the discrepancy lies between $0.4\%$  ($Tg=0.5$ hrs) and $2.4\%$ ($Tg=4$ hrs) in the range $f\in[10^{-5},10^{-3}]$. The Hann windowing is applied to reduce the noise effect due to the segmentation, i.e. jumps in the values at the segments boundaries resulting in $\approx 1/f$ noise (see panel {\bf b}, pink curve). 

The other three techniques allow  recovery of the full-frequency spectra, see panels ({c},{d}) of figure \ref{fig:method}. To all these spectra a light smoothing has been applied  homogeneously over the entire spectral range using a 9-point running average. The computation of the two-point correlation function is highly affected by the sparsity of data. As a result, the spectrum of the correlation function computed from the original gapped data 
(orange curves) is not physically significant. The peaks are entirely due to the gap distribution; causality is shown by the comparison with the Fourier transform of a boolean characteristic function $\Phi$, which is zero-valued where the data are missing (cyan curves). The correlation function computed from linearly interpolated data in the full-period leads to much better convergence, leading to the spectra represented in pink. The maximum likelihood reconstruction is a non-deterministic recovery, even though it is constrained by the true data where these are available. A complete description of the technique is given by \cite{rybicki1992a} and an application can by found in \cite{rybicki1992b}. It requires an estimation of the two-point correlation function and it also allows one to account for noise in data. We used the same correlations computed for the previous method. In this case the size of filled gaps $Tg$ is a parameter to be chosen, and here the goodness of the correlation function allows us to recover the full sequence (green spectra).

Compressed sensing (CS in the following, \cite{donoho2006,candes2006}) is a recent theory that provides guarantees for the  reconstruction of (exactly or approximately) sparse signals, namely signals with many null (or approximately null) components, from linear, compressed measurements. In mathematical terms, CS studies the underdetermined linear system $A\mathbf{y}=\mathbf{x}$, where $A$ is matrix of size $m\times n$ with $m< n$, and $\mathbf{y}$ and $\mathbf{x}$ have consistent dimensions. The available data vector $\mathbf{x}$ is then a linear compression of the unknown $\mathbf{y}$, which is assumed to be sparse. CS theory provides conditions that make such problems well posed, that is, with a unique solution. In particular, much effort has been devoted to study which families of sensing matrices $A$ guarantee the possibility of recovery. Among these, partial Fourier matrices (say, discrete Fourier transform matrices with missing rows) have been recently studied (see, e.g., \cite{rud06, dua11, xu15}),
 motivated in particular by the applications in medical imaging problems such as MRI \cite{lus08}. In the mentioned works, theoretical 
guarantees on partial Fourier matrices for CS are provided, in terms of the number of necessary measurements and positioning of the missing rows.
 The problem of spectrum recovery from missing data can be interpreted as an undetermined linear system  $A\mathbf{y}=\mathbf{x}$ where $A$ is a partial Fourier matrix, $\mathbf{x}$ is the vector of lacunous data, and  $\mathbf{y}$ is the spectrum to recover. 
Such a spectrum can be considered as an approximation of the spectrum of  the physical signal and  will have a  level of sparsity increasing proportionally to the number of gaps in the data. This fact can be accepted since the density of information in the time series is sufficient to allow a quantitative estimate of the power spectral distribution and in particular of the related power decay in the range of frequencies recovered. 
 The CS approach allows to perform spectral analysis with no previous interpolation of the lacunous data. On the other hand, once the spectrum $\mathbf{y}$ is estimated, one can perform an inversion of the Fourier transform and obtain an estimate of the data without gaps.
 CS provides the theoretical tools to tackle our recovery problem. Moreover, it is also feasible in practice, even when data sets are large. Specifically, in this work we formulated the problem as a Basis Pursuit and obtained the numerical solution through the  SPGL1 Matlab solver for sparse problems (see \cite{spgl07} for theoretical and practical details).  SPGL1 is suitable for the Fourier framework, as it deals with complex variables. Moreover, it allows us to cope with data of large dimension, since the sensing matrix $A$ can be defined as function instead of explicitly storing the whole matrix.

 The relative error in the spectral slope, with respect to the true value (black curves) for these three methods is below $2.4\%$, in the range $f\in[10^{-6},5\cdot10^{-3}]$ of \textit{Synt1}, and $f\in[10^{-6},10^{-3}]$ of \textit{Synt2}.




\subsection{Plasma and magnetic field spectra}

\begin{figure*}
 	\centering
 		\includegraphics[width=0.85\textwidth]{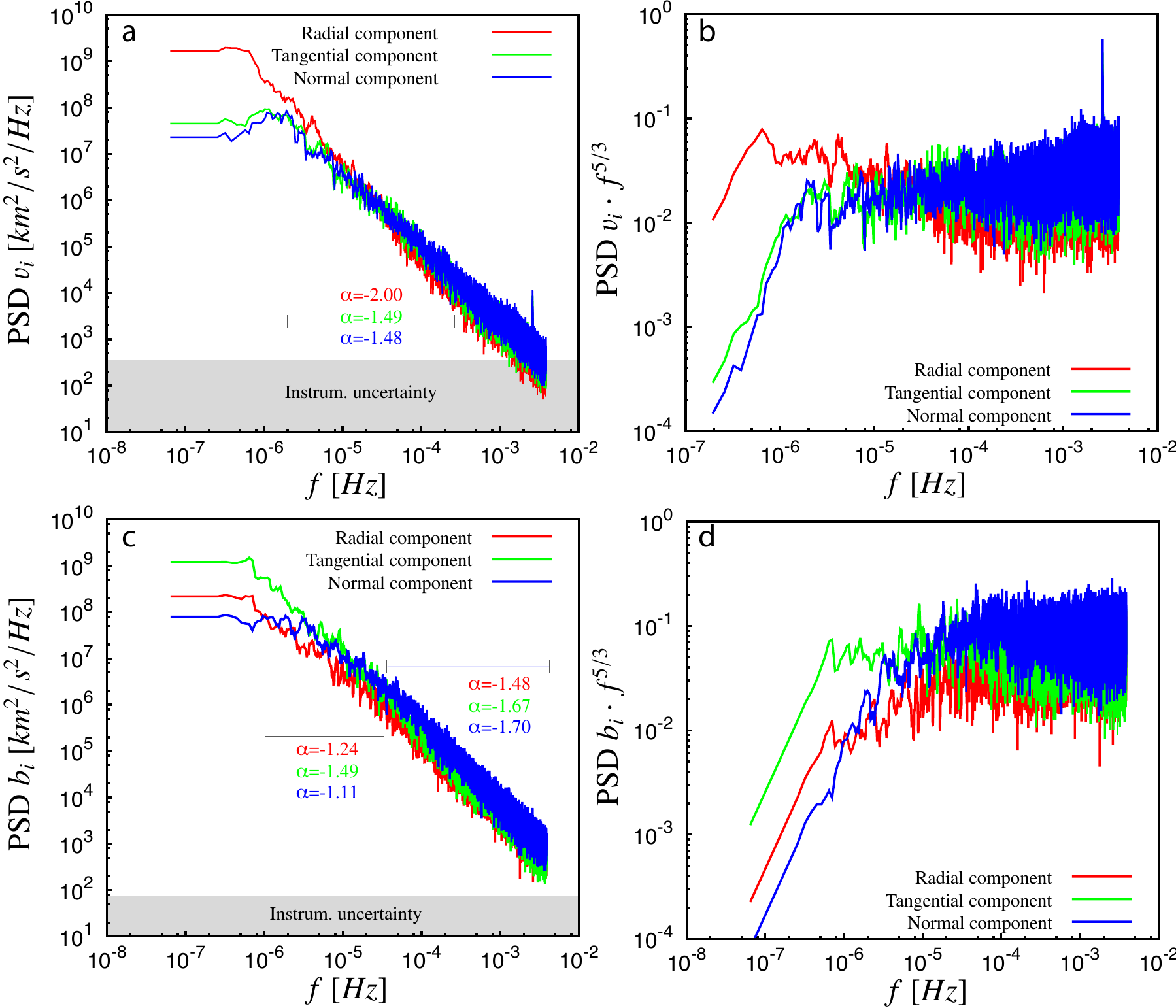} 		
 		\caption{Compensated spectra of the components of plasma velocity (top panel) and magnetic field (bottom panel) computed by using the correlations method. The exponents used for compensation are the ones of the high frequency range for the module spectra, as reported in table \ref{tab:data_exponent}. In this way, the presence of a slope in the compensated spectrum of each component indicates the anisotropy of the field. In panels (\textit{b,d}), the compensated spectra of kinetic and magnetic energies is represented including periods from 3 minutes to 180 days. The changes of exponent laws and the differences at high frequencies are here more visible.}
	\label{fig:component} 
 \end{figure*}

The spectra obtained for each component of magnetic field and plasma velocity are represented in figure \ref{fig:component}.
It should be noted that all the gap recovery methods converge at similar exponent, as shown in Fig S 1  of  Supplementary Information.   

\newpage
             
Looking at the power spectra of each component of the velocity field, shown in the top panels of figure \ref{fig:component} and resumed in table \ref{tab:data_exponent} (\textit{a}), the behavior of the system at low frequencies ($f<4\cdot10^{-4}$) shows that the exponent is steeper for the radial component rather than the other two. In particular, the spectral index in the radial direction is $\alpha_{v_R}\approx-2.00$. Considering that these frequency spectra can be interpreted as perpendicular-wavenumber spectra (see eq. \ref{eq:perp}), a slope of about 2 for the radial component can identify an inertial range of anisotropic weak MHD turbulence \citep{galtier2000} with relevant effects of parallel sweeping (interaction between coherent structures of different scales, see \cite{zhou2004}).
At high frequencies, the situation is reversed and the spectral index of the radial component is $\alpha_{v_R}\approx-1.18$, smaller than the other two ($\alpha_{v_T}\approx-1.28$, $\alpha_{v_N}\approx-1.48$).
Considering the power spectrum of the kinetic energy, the exponent found is $\alpha_{E_k}\approx-1.67$ in the low frequency domain. Such a value is consistent with the Kolmogorov theory,  which predicts an exponent of -5/3 (see also \citet{marsch89}).  In the high frequency range, for $f\in[3\cdot10^{-4},2\cdot10^{-3}]$  the fit gives the value of $\alpha_{E_k}\approx-1.33$.
As pointed out by \cite{matt89} and reviewed by \cite{zhou2004},  energy spectra exponents ranging from $-5/3$ to $-3/2$ can indicate the presence of Alfv\'enic waves.
For high frequencies, we can consider that the Iroshnikov-Kraichnan MHD cascade model (spectral index of -3/2) indicates a quasi 2D behavior of the field. As shown by \cite{matt89}, lower exponents (as the $\alpha_{E_k}\approx-1.26$ such as those found for $f>4\cdot10^{-4}$) are typical of weak turbulence, in which the field is characterized by strong mean values and relatively small fluctuations and in which nonlinear interactions are not relevant. The flattening in the velocity spectra at high frequencies has been observed by other authors \cite{matthaeus1982b, roberts2010}. The first paper pointed out that the flattening may be due to aliasing, but the second paper excluded this hypothesis. This flattening was also found for the proton density and temperature fluctuations \citep{marsch1990b}, as well as for the Els\"{a}sser variables \citep{marsch1990a}. It is most typical of high-speed streams below 1 AU, in these cases the frequency range is more extended and the change of slope occurs at  $f\approx10^{-7}$. 
Here we cannot exclude that such evident flattening in the velocity spectra be partially due to the level of noise in the data. The instrumental uncertainty on the velocity components is about $\pm 2$ km/s. Modeling this as a uniform noise of amplitude  $\pm 2$ km/s, we observe from the synthetic reference that this could influence the last spectral decade (see figure \ref{fig:generale}).
The magnetic field components present a very different phenomenology with respect to the plasma velocity: for each component the spectra have higher exponents in the high frequency range ($f>3\cdot10^{-5}$) and they tend to become flatter at low frequencies, as shown in figure \ref{fig:component} (\textit{c-d}) and in tables \ref{tab:data_exponent} (\textit{b-c}); moreover, an anisotropic behavior is also observed both at high and low frequencies: in particular, the radial component is always lower than the tangential one. The normal component, instead, has a spectral index analogous to the radial one a low frequencies (for $f<3\cdot10^{-5}$, considering Alfv\'en units, $\alpha_{b_R}\approx-1.24$ and $\alpha_{b_N}\approx-1.11$, when $\alpha_{b_T}\approx-1.49$), while it becomes similar to the tangential exponent at high frequencies (for $f>3\cdot10^{-5}$, considering Alfv\'en units, $\alpha_{b_T}\approx-1.67$ and $\alpha_{b_N}\approx-1.70$, when $\alpha_{b_T}\approx-1.48$).

The spectral exponent for the magnetic energy is $\alpha_{E_m}\approx-1.65$ at high frequencies, as predicted by the Kolmogorov law. At low frequencies the exponents drop to values around $\alpha_{E_m}\approx1.34$, thus the energy cascade is very damped in this frequency range.
Though the magnetic field fluctuations in the inertial
range often follow a Kolmogorov -5/3 behavior \cite{podesta2007}, spectral indices around 1.8 have been recently observed at about 1 AU by other authors as \cite{safrankova2013}.
Spectra presented here are in good agreement with those found by \cite{matthaeus1992a} (magnetic spectral index of -1.7 at high frequencies for Voyager 1 data at about 5 AU) and \cite{klein} (magnetic spectral index of -1.17 at low frequencies and -1.88 at high frequencies for Voyager 1 data at 4 AU), where an analysis for V1 spectra of magnetic modules respectively at 5 and 10 AU can be found for a frequency range from $10^{−7}\div10^{−4}$.
\begin{table}
	\caption{\label{tab:data_exponent} Synthesis of exponents found for plasma velocity and  magnetic field (both in tesla and Alfv\'en units), highlighting the different frequency ranges. Maximum interpolation error is of the order of $0.07$.}
	\centering
	\begin{tabular}{cllll}
		\hline\\
		$f$ range &  ${v}_R$ & ${v}_T$ & ${v}_N$ &  $|\mathbf{v}|$\\
		\hline\\
		$\hphantom{1\cdot\ }10^{-6} \div 4\cdot10^{-4}$
		& -2.00 & -1.49  & -1.48 & -1.67 \\	
		$4\cdot10^{-4} \div 5\cdot10^{-3}$
		& -1.18 & -1.26 & -1.48 & -1.33 \\
		\hline
	\end{tabular}\vspace{1em}
	\begin{tabular}{cllll}
		\hline\\
		$f$ range &  ${B}_R$ & ${B}_T$ & ${B}_N$ &  $|\mathbf{B}|$\\
		\hline\\
		$\hphantom{1\cdot\ }10^{-6} \div 3\cdot10^{-5}$
		& -1.06 & -1.46  & -0.85 & -1.21 \\
		$3\cdot10^{-5} \div 5\cdot10^{-3}$
		& -1.56 & -1.72 & -1.77 & -1.72 \\
		\hline
	\end{tabular}\vspace{1em}
	\begin{tabular}{cllll}
		\hline\\
		$f$ range &  ${b}_R$ & ${b}_T$ & ${b}_N$ &  $|\mathbf{b}|$\\
		\hline\\
		$\hphantom{1\c	dot\ }10^{-6} \div 3\cdot10^{-5}$
		& -1.24 & -1.49  & -1.11 & -1.34 \\
		$3\cdot10^{-5} \div 5\cdot10^{-3}$
		& -1.48 & -1.67 & -1.70 & -1.65 \\
		\hline
	\end{tabular} 
\end{table}
	\begin{table}
	\caption{\label{length}Turbulence temporal scales. $T$ represents the integral scale, while $\tau$ the Taylor micro-scale. A study dedicated to helicities in the same period can be found in \cite{iovieno2015}}
	\centering
			\begin{tabular}{@{}llll}
				\hline
				\multicolumn{2}{c}{Parameter} &  \multicolumn{2}{c}{Value}                      \\
				\hline
				\boldmath$T_{E_v}$          & Kinetic correlation length           & $26.8$  & days           \\
				\boldmath$T_{E_m}$          & Magnetic correlation length          & $25.9$  & days           \\
				\boldmath$T_{N}$          & Density correlation length    & $13.1$  & days           \\
				\boldmath$T_{vth}$          & Thermal speed correlation length & $11.0$  & days           \\
				\boldmath$T_{Hc}$          & Cross-helicity correlation length    & $25.1$  & days           \\
				\boldmath${\tau_v}$    & Kinetic Taylor micro-scale                 & $0.85$  & hrs            \\
				\boldmath${\tau_m}$    & Magnetic Taylor micro-scale                & $0.85$  & hrs           \\
				\boldmath${\tau_N}$    & Density Taylor micro-scale                & $1.04$  & hrs           \\
				\boldmath${\tau_{vth}}$    & Thermal speed Taylor micro-scale                & $0.85$  & hrs           \\
				\boldmath${\tau_{Hc}}$    & Cross-helicity Taylor micro-scale                & $5.17$  & hrs           \\
				\hline
			\end{tabular}
	\end{table}

\begin{figure}
	\includegraphics[width=.45\textwidth]{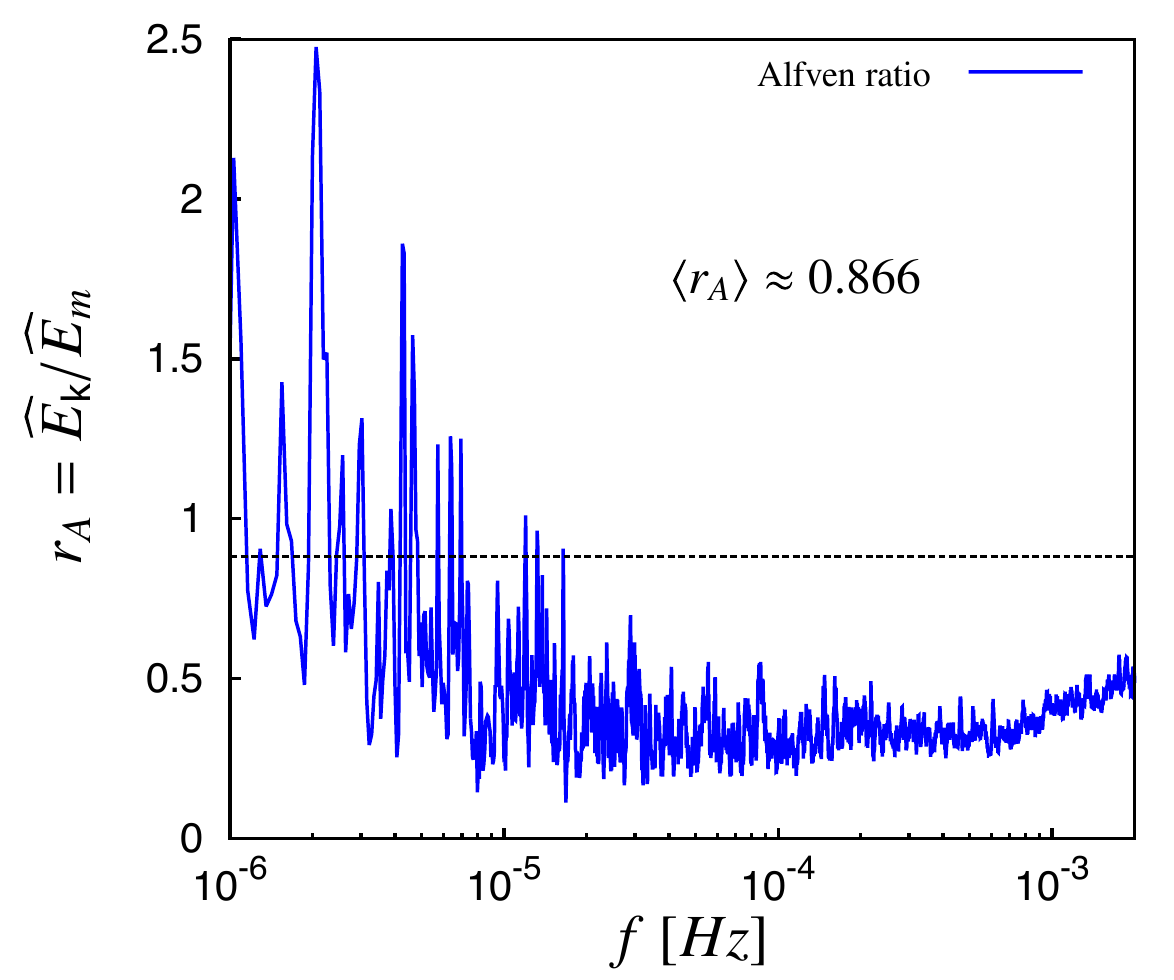}
	\caption{Alfven ratio $\widehat{r}_A$ in semilog plot. Differences between slopes can be observed looking at the curve trend. $r_A$ reaches values lower than $0.5$, as already seen for slow solar wind (\textit{Marsch and Tu, J. Geophys. Res.,95, 8211, 1990})}
	\label{fig:alf_ratio}
\end{figure}

The different behavior of the power spectra of kinetic and magnetic energies can be appreciated by considering the Alfv\`en ratio, defined as
\begin{equation}
\widehat{r}_A(f) =\frac{\widehat{E}_v(f)}{\widehat{E}_b(f)}
\end{equation}
and represented in Figure \ref{fig:alf_ratio}.

\begin{figure}
	\includegraphics[width=.45\textwidth]{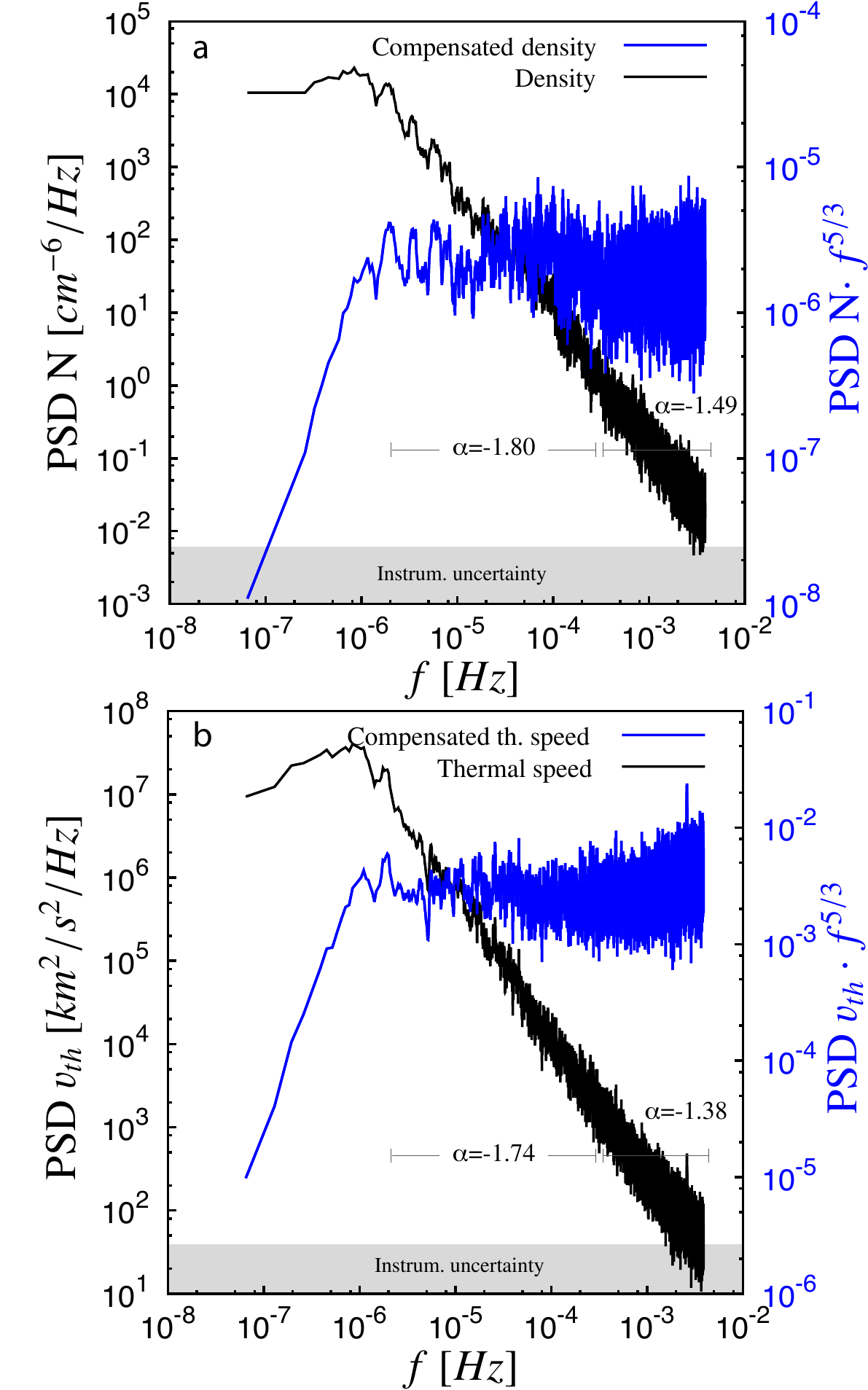}
	\caption{Density and thermal speed spectra} 
		\label{fig:den-temp-spectra}
\end{figure}

The Alfv\'en ratio is usually less than unity in the inertial range \cite{tu1995} \cite{matthaeus1982b}. The change occurs when $\widehat{r}_A$ reaches its minimum values. Moreover, the minimum value is lower than 0.5, as observed for slow solar wind inside 1 AU \citep{orlando1997}. In general, leaving aside the frequency of the minimum which depends on the distance from the Sun, similar evolution of $\widehat{r}_A$ with frequency can be observed \cite{marsch1990a}. For completeness, in Figure \ref{fig:den-temp-spectra}, spectra of the density and thermal speed are reported. The behavior is similar for the integral scale and inertial decay in the central part. A light steepening of the inertial decay is observed for the thermal speed analogous to that of the plasma normal component.

\section{Conclusions }
In this work we computed the power spectra of the solar slow wind, density, thermal speed and magnetic field at 5 AU at low latitude from Voyager 2 measurements. By using data reconstruction techniques to overcome the missing data issues, we have been able to determine the spectra for a frequency range extending over five decades ($10^{-7}-10^{-2}$ Hz). This extended range, much wider than in any other previous  study at this distance, allows us to observe the changes in the spectral slopes providing information on the structure of the solar wind.
The analysis procedures have been validated by testing three different data recovery methodsr, linear interpolation, two-point correlations, and maximum likelihood recovery and compressed sensing, on synthetic data which mimic the behavior of the well-known homogeneous and isotropic turbulence system, which have been made lacunous by projecting the same gap distribution as in the V2 plasma data. 
The plasma velocity spectrum presents an inertial range with an exponent close to the Kolmogorov model up to $f=3\cdot 10^{-4}$ Hz, while some flattening occurs at higher frequencies. The magnetic spectrum shows a change of spectral index at about $f=5\cdot 10^{-5}$ Hz, in agreement with Ulysses data near the ecliptic at 4.8 AU. Above this frequency the slope is $-1.76\pm0.06$ and remains constant in the whole range of observed frequencies. We can therefore conclude that the inertial range extends at least from $f=5\cdot10^{-5}$ to $f=5\cdot 10^{-3}$ Hz. 
Given the direction of the mean magnetic field, these spectra can be seen as spatial spectra in the perpendicular direction to the mean magnetic field when the Taylor frozen-flow assumption is used. The variation of the spectral index can be due to the presence of Alfven waves or to anisotropies with relevant effects of the parallel sweeping due to the large scale magnetic fluctuations. In fact, the anisotropy appears to be significant at frequencies below $f= 10^{-5}$ Hz, when most of the energy tends to be concentrated into the radial component of the velocity fluctuations and in the tangential component of the magnetic field fluctuations.  Moreover, the Alfven ratio, which remains small for most of the frequency range, becomes larger than 0.5  at the lowest frequencies, below $f=10^{-5}$ Hz. At high frequencies there is a marked dominance of magnetic energy with respect to kinetic energy. 

For all the gap compensation methods here presented, practical details concerning software issues are described in the Supplementary Information.

\begin{acknowledgments}

This collaboration between the Politecnico di Torino and M.I.T. was sponsored by the Progetto MITOR with support by the Fondazione Compagnia Di San Paolo. JDR was supported by NASA under the Voyager project. 

\end{acknowledgments}


\bibliographystyle{agufull08}
\bibliography{JGR_bibliography_new}

\begin{thebibliography}{30}
\providecommand{\natexlab}[1]{#1}
\expandafter\ifx\csname urlstyle\endcsname\relax
  \providecommand{\doi}[1]{doi:\discretionary{}{}{}#1}\else
  \providecommand{\doi}{doi:\discretionary{}{}{}\begingroup
  \urlstyle{rm}\Url}\fi

\bibitem[{\textit{Berg and Friedlander}(2007)}]{spgl07}
Berg, E. V.~D., and M.~P. Friedlander (2007), {SPGL1}: A solver for large-scale
  sparse reconstruction, http://www.cs.ubc.ca/labs/scl/spgl1.

\bibitem[{\textit{Bridge et~al.}(1977)\textit{Bridge, Belcher, Butler, Lazarus,
  Mavretic, Sullivan, Siscoe, and Vasyliunas}}]{bridge1977}
Bridge, H.~S., J.~W. Belcher, R.~J. Butler, A.~J. Lazarus, A.~M. Mavretic,
  J.~D. Sullivan, G.~L. Siscoe, and V.~M. Vasyliunas (1977), The plasma
  experiment on the 1977 voyager mission, \textit{Space Sci. Rev.},
  \textit{21}, 259--287.

\bibitem[{\textit{Bruno and Carbone}(2013)}]{bruno2013}
Bruno, R., and V.~Carbone (2013), The the solar wind as a turbulence
  laboratory, \textit{Living Rev. Solar Phys.}, \textit{10}(2).

\bibitem[{\textit{Cand{\'e}s et~al.}(2006)\textit{Cand{\'e}s, Rombreg, and
  Tao}}]{candes2006}
Cand{\'e}s, E., J.~Rombreg, and T.~Tao (2006), Robust uncertainty principles:
  exact signal reconstruction from highly incomplete frequency information,
  \textit{IEEE Trans. on Information Theory}, \textit{52}, 489--509.

\bibitem[{\textit{Donoho}(2006)}]{donoho2006}
Donoho, D.~L. (2006), Compressed sensing, \textit{IEEE T. Inform. Theory},
  \textit{52}, 1289--1306.

\bibitem[{\textit{Duarte and Eldar}(2011)}]{dua11}
Duarte, M., and Y.~Eldar (2011), Structured compressed sensing: From theory to
  applications, \textit{Signal Processing, IEEE Transactions on},
  \textit{59}(9), 4053--4085, \doi{10.1109/TSP.2011.2161982}.

\bibitem[{\textit{Galtier et~al.}(2000)\textit{Galtier, Nazarenko, Newell, and
  Pouquet}}]{galtier2000}
Galtier, S., S.~Nazarenko, A.~Newell, and A.~Pouquet (2000), A weak turbulence
  theory for incompressible magnetohydrodynamics, \textit{J. Plasma Phys.},
  \textit{63}(5), 447--488, \doi{10.1017/S0022377899008284}.

\bibitem[{\textit{Horbury and Tsurutani}(2001)}]{horbury2001}
Horbury, T.~S., and B.~T. Tsurutani (2001), \textit{The heliosphere near solar
  minimum: the Ulysses perspective}, chap. Ulysses measurements of turbulence,
  waves and discontinuities, Springer, in association with Praxis Publishing.

\bibitem[{\textit{Iovieno et~al.}(2015)\textit{Iovieno, Gallana, Fraternale,
  Richardson, Opher, and Tordella}}]{iovieno2015}
Iovieno, M., L.~Gallana, F.~Fraternale, J.~D. Richardson, M.~Opher, and
  D.~Tordella (2015), Cross and magnetic helicity in the outer heliosphere from
  voyager 2 observations, \textit{European Journal of Mechanics B/Fluids}.

\bibitem[{\textit{Klein et~al.}(1991)\textit{Klein, Matthaeus, Roberts, and
  Goldstein}}]{klein}
Klein, L.~W., W.~H. Matthaeus, D.~A. Roberts, and M.~L. Goldstein (1991), Solar
  wind seven, \textit{J. Plasma Phys.}, \textit{7}, 197--200.

\bibitem[{\textit{Lustig et~al.}(2008)\textit{Lustig, Donoho, Santos, and
  Pauly}}]{lus08}
Lustig, M., D.~Donoho, J.~Santos, and J.~Pauly (2008), Compressed sensing mri,
  \textit{Signal Processing Magazine, IEEE}, \textit{25}(2), 72--82,
  \doi{10.1109/MSP.2007.914728}.

\bibitem[{\textit{Marsch and Tu}(1989)}]{marsch89}
Marsch, E., and C.~Tu (1989), Dynamics of correlation-functions with
  els{\"a}sser variables for inhomogeneous mhd turbulence, \textit{J. Plasma
  Phys.}, \textit{41}, 479--491.

\bibitem[{\textit{Marsch and Tu}(1990{\natexlab{a}})}]{marsch1990a}
Marsch, E., and C.-Y. Tu (1990{\natexlab{a}}), On the radial evolution of mhd
  turbulence in the inner heliosphere, \textit{J. Geophys. Res.},
  \textit{95}(14), 8211--8229.

\bibitem[{\textit{Marsch and Tu}(1990{\natexlab{b}})}]{marsch1990b}
Marsch, E., and C.-Y. Tu (1990{\natexlab{b}}), Spectral and spatial evolution
  of mhd turbulence in the inner heliosphere, \textit{J. Geophys. Res.},
  \textit{95}(14), 11,945--11,956.

\bibitem[{\textit{Matthaeus and Goldstein}(1982)}]{matthaeus1982b}
Matthaeus, W.~H., and M.~L. Goldstein (1982), Measurement of the rugged
  invariants of magnetohydrodynamic turbulence in the solar wind, \textit{J.
  Geophys. Res.}, \textit{87}(16), 6011--6028.

\bibitem[{\textit{Matthaeus and Goldstein}(1992)}]{matthaeus1992a}
Matthaeus, W.~H., and M.~L. Goldstein (1992), Evaluation of magnetic helicity
  in homogeneous turbulence, \textit{Phys. Rev. Letters}, \textit{48}(18).

\bibitem[{\textit{Matthaeus and Zhou}(1989)}]{matt89}
Matthaeus, W.~H., and Y.~Zhou (1989), Extended inertial range phenomenology of
  magnetohydrodynamic turbulence, \textit{Phys. Fluids B}, \textit{1}, 1929.

\bibitem[{\textit{McComas et~al.}(2003)\textit{McComas, Bame, and
  et~al.}}]{mccomas1998}
McComas, D.~J., S.~J. Bame, and B.~L.~B. et~al. (2003), Ulysses' return to the
  slow solar wind, \textit{Geophys. Res. Lett.}, \textit{71}, 3315--3325.

\bibitem[{\textit{Monin and Yaglom}(1971)}]{monin_book}
Monin, A.~S., and A.~M. Yaglom (1971), \textit{Statistical fluid mechanics,
  Volume 2}, vol.~II, MIT Press.

\bibitem[{\textit{Montgomery et~al.}(1987)\textit{Montgomery, Brown, and
  Matthaeus}}]{montgomery1987}
Montgomery, D., M.~Brown, and W.~Matthaeus (1987), Density fluctuation spectra
  in magnetohydrodynamic turbulence, \textit{J. Geophys. Res.},
  \textit{92}(A1), 282--284, \doi{10.1029/JA092iA01p00282}.

\bibitem[{\textit{Orlando et~al.}(1997)\textit{Orlando, Lou, Peres, and
  Rosner}}]{orlando1997}
Orlando, S., Y.~Q. Lou, G.~Peres, and R.~Rosner (1997), Alfvenic fluctuations
  in fast and slow solar winds, \textit{J. Geoph. Res.}, \textit{102},
  24,139--24,149.

\bibitem[{\textit{Podesta et~al.}(2007)\textit{Podesta, Roberts, and
  Goldstein}}]{podesta2007}
Podesta, J.~J., D.~A. Roberts, and M.~L. Goldstein (2007), Spectral exponents
  of kinetic and magnetic energy spectra in solar wind turbulence,
  \textit{Astrophys J.}, \textit{664}, 543--548.

\bibitem[{\textit{Press and Rybicki}(1992)}]{rybicki1992b}
Press, W.~H., and G.~B. Rybicki (1992), The time delay of gravitational lens
  0957+561. i. methodology and analysis of optical photometric data,
  \textit{The Astroph. J.}, \textit{385}, 404--415.

\bibitem[{\textit{Roberts}(2010)}]{roberts2010}
Roberts, D.~A. (2010), Evolution of the spectrum of solar wind velocity
  fluctuations from 0.3 to 5 au, \textit{J. Geophys. Res.}, \textit{115},
  A12,101.

\bibitem[{\textit{Rudelson and Vershynin}(2006)}]{rud06}
Rudelson, M., and R.~Vershynin (2006), Sparse reconstruction by convex
  relaxation: Fourier and gaussian measurements, in \textit{Information
  Sciences and Systems, 2006 40th Annual Conference on}, pp. 207--212,
  \doi{10.1109/CISS.2006.286463}.

\bibitem[{\textit{Rybicki and Press}(1992)}]{rybicki1992a}
Rybicki, G.~B., and W.~H. Press (1992), Interpolation, realization, and
  reconstruction of noisy, irregularly sampled data, \textit{The Astroph. J.},
  \textit{398}, 169--176.

\bibitem[{\textit{Tu and Marsch}(1995)}]{tu1995}
Tu, C.-Y., and E.~Marsch (1995), Mhd structures, waves and turbulence in the
  solar wind: observations and theories, \textit{Space Sci. Rev.}, \textit{73},
  1--210.

\bibitem[{\textit{\v{S}afr\'{a}nkov\'{a}
  et~al.}(2013)\textit{\v{S}afr\'{a}nkov\'{a}, N\v{e}me\v{c}ek, P\v{r}ech, and
  Zastenker}}]{safrankova2013}
\v{S}afr\'{a}nkov\'{a}, J., Z.~N\v{e}me\v{c}ek, L.~P\v{r}ech, and G.~N.
  Zastenker (2013), Ion kinetic scale in the solar wind observed, \textit{Phys.
  Rev. Letters}, \textit{110}, 025,004.

\bibitem[{\textit{Xu and Xu}(2015)}]{xu15}
Xu, G., and Z.~Xu (2015), Compressed sensing matrices from fourier matrices,
  \textit{IEEE Trans. Inf. Theory}, \textit{61}(1), 469--478,
  \doi{10.1109/TIT.2014.2375259}.

\bibitem[{\textit{Zhou et~al.}(2004)\textit{Zhou, Matthaeus, and
  Dmitruk}}]{zhou2004}
Zhou, Y., W.~Matthaeus, and P.~Dmitruk (2004), Colloquium: Magnetohydrodynamic
  turbulence and time scales in astrophysical and space plasmas, \textit{Rev.
  Mod. Phys.}, \textit{76}(4), 1015--1035, \doi{10.1103/RevModPhys.76.1015}.

\end{thebibliography}

\end{article}

\end{document}